\documentclass[prd,twocolumn,nofootinbib,longbibliography]{revtex4-1}
\usepackage{graphicx}
\usepackage{amssymb}
\usepackage{amsmath}
\usepackage{slashed}
\usepackage{hyperref}

\newcommand{\abs}[1]{|#1|}

\usepackage{feynmp}
\def\lsim{\mathrel{\raise.3ex\hbox{$<$\kern-.75em\lower1ex\hbox{$\sim$}}}}
\def\gsim{\mathrel{\raise.3ex\hbox{$>$\kern-.75em\lower1ex\hbox{$\sim$}}}}

\usepackage{color}

\begin{document}
\begin{fmffile}{diagrams}

\title{Modifying dark matter indirect detection signals by thermal effects at freeze-out}

\author{Andi Hektor}
\email[]{andi.hektor@cern.ch}
\affiliation{National Institute of Chemical Physics and Biophysics, R\"avala 10, 10143 Tallinn, Estonia}
\author{Kristjan Kannike}
\email[]{kristjan.kannike@cern.ch}
\affiliation{National Institute of Chemical Physics and Biophysics, R\"avala 10, 10143 Tallinn, Estonia}
\author{Ville Vaskonen}
\email[]{ville.vaskonen@kbfi.ee}
\affiliation{National Institute of Chemical Physics and Biophysics, R\"avala 10, 10143 Tallinn, Estonia}

\date{\today}

\begin{abstract}
We present an extension of the Standard Model, containing a fermion dark matter candidate and two real scalar singlets, where the observed dark matter abundance is produced via freeze-out before the electroweak phase transition. We show that in this case the dark matter annihilation channels determining its freeze-out are different from those producing indirect detection signal. We present a benchmark model where the indirect annihilation cross-section differs from the freeze-out one. The model also has a gravitational wave signature due to the first order electroweak phase transition.
\end{abstract}

\maketitle

\section{Introduction}

The existence of dark matter (DM) is among the few indications of physics beyond the Standard Model (SM). In recent years, many popular models of weakly interacting massive particles (WIMPs) have become under pressure~\cite{Escudero:2016gzx}, as the bounds from direct detection, e.g. the LUX~\cite{Akerib:2016vxi}, PandaX~\cite{Tan:2016zwf} and XENON1T~\cite{Aprile:2017iyp} detectors, are getting very stringent. While these constraints can be evaded, for example, by `secluded' WIMPs~\cite{Pospelov:2007mp,Arcadi:2016qoz}, semi-annihilation~\cite{DEramo:2010keq, Hambye:2008bq, Hambye:2009fg, Arina:2009uq, Belanger:2012vp, Belanger:2012zr, Belanger:2014bga, Aoki:2014cja, Cai:2015zza, Bhattacharya:2017fid} or a pseudoscalar mediator~\cite{Boehm:2014hva, Hektor:2014kga, Kainulainen:2015sva, Hektor:2015zba, Hektor:2017ftg}, it is usual for WIMPs that the same processes which determine DM freeze-out usually also cause an indirect detection signal.

Constraints from indirect detection are getting stronger as well, mainly due to $\gamma$-ray measurements from the Galactic Centre and dwarf spheroidal satellite galaxies~\cite{Charles:2016pgz, Ahnen:2016qkx}, and a possible future detection (e.g. by the CTA experiment~\cite{Pierre:2014tra}) can exclude the WIMP paradigm. Well-known examples to decouple the freeze-out cross-section from the indirect one are the Sommerfeld enhancement (e.g. \cite{Hisano:2004ds,Feng:2010zp}) or the Breit-Wigner resonance (e.g. \cite{Guo:2009aj}). Moreover, in the feebly interacting massive particle (FIMP) scenario (e.g.~\cite{Bernal:2017kxu}), if the DM abundance is determined by freeze-out in a hidden sector, one can obtain an indirect detection cross-section very different from the usual WIMP case, because the hidden sector temperature differs from the visible sector one at the freeze-out~\cite{Feng:2008mu} (for a recent development see \cite{Heikinheimo:2018}).

Besides changing average relative velocity, other properties of DM and DM-SM mediators could be modified by the thermal evolution of the Universe. Thermal effects may modify masses and interactions of particles due to the existence of several minima of the potential. For example, at the time of its freeze-out, the mass of the DM particle can be different than at low temperatures, thus changing the interactions of DM with the SM particles compared to the usual WIMP~\cite{Rychkov:2007uq, Cohen:2008nb, Strumia:2010aa}. Furthermore, there can be overproduction of DM via freeze-out, which is later corrected by partial decay of the DM in a phase where the WIMP is not stable before the electroweak phase transition (EWPT)~\cite{Baker:2016xzo,Kobakhidze:2017ini}. Thermal changes of DM properties and evolution may include also `forbidden' annihilation channels~\cite{DAgnolo:2015ujb, Griest:1990kh}, cannibal DM~\cite{Pappadopulo:2016pkp}, and dynamic freeze-in~\cite{Baker:2017zwx}. 

The purpose of the current work is to modify the WIMP scenario to decouple the annihilation cross-section of indirect detection from the freeze-out cross-section.
 The thermal evolution of the Universe can temporarily open new WIMP annihilation channels. To exploit that effect, we present a scenario where the DM freeze-out occurs before the EWPT. We extend the SM by two real scalar singlets, and a singlet fermion as the DM candidate. The thermal evolution of the Universe proceeds in a two-step manner~\cite{Espinosa:2011ax, Espinosa:2011eu, Cline:2012hg, Alanne:2014bra, Alanne:2016wtx, Tenkanen:2016idg, Vaskonen:2016yiu}. First, at some high temperature there is a transition from zero field values to a minimum where the lighter singlet scalar gets a non-zero vacuum expectation value (VEV). Second, the transition to the usual EW vacuum follows, where the singlet VEVs are zero. 

The non-zero singlet VEV opens efficient DM annihilation channels that yield the observed DM relic density, but which are closed in the EW-breaking minimum. This separates the DM annihilation processes, which determine its freeze-out, from the indirect detection signal, corresponding to the DM annihilation processes in the EW-breaking minimum. Na\"ive calculation assuming the EW minimum throughout the evolution of the Universe gives a wrong result.

The EWPT in this model is typically of first order, thus generating a stochastic gravitational wave (GW) background~\cite{Steinhardt:1981ct, Hogan:1984hx, Witten:1984rs}. This can potentially be probed in future space based GW interferometers~\cite{Corbin:2005ny, Seoane:2013qna}, as has previously been studied in several extensions of the SM~\cite{Huber:2015znp, Kakizaki:2015wua, Leitao:2015fmj, Artymowski:2016tme, Huang:2016cjm, Huang:2016odd, Vaskonen:2016yiu, Chala:2016ykx, Dorsch:2016nrg, Hashino:2016xoj, Kurup:2017dzf, Beniwal:2017eik, Marzola:2017jzl,  Kang:2017mkl, Kobakhidze:2017mru, Baldes:2017rcu, Huang:2017rzf, Chao:2017vrq, Akula:2017yfr, Demidov:2017lzf}.

The paper is organized as follows: in Section~\ref{sec:model} we present the model and discuss theoretical and phenomenological restrictions on the parameter space. In Section~\ref{sec:ewpt} we describe thermal effects and the EWPT. DM freeze-out is treated in Section~\ref{sec:freezeout}, the indirect detection signal in Section~\ref{sec:detection}, and the the GW background in Section~\ref{sec:gw}. We summarize our key conclusions in Section~\ref{sec:conclusions}.

\section{The model}
\label{sec:model}

We consider an extension of the SM which comprises a Dirac fermion DM candidate $\chi$ and two real scalar singlets $S_{1}$ and $S_{2}$. The Lagrangian thus includes the terms\footnote{We consider pseudoscalar Yukawa couplings to avoid velocity suppression in the DM indirect detection cross section that arises for scalar Yukawa couplings.}
\begin{equation}
\begin{split}
  \mathcal{L} &\supset \bar{\chi} \slashed{\partial} \chi + \abs{D_{\mu} H}^{2}
  + \frac{(\partial_{\mu} S_{1})^{2}}{2} + \frac{(\partial_{\mu} S_{2})^{2}}{2} - M_{\chi} \bar{\chi} \chi \\ 
  &- i y_{1} S_{1} \bar{\chi} \gamma_5 \chi - i y_{2} S_{2} \bar{\chi} \gamma_5 \chi - V(H, S_{1},S_{2}),
\end{split}
\end{equation}
where $H$ is the SM Higgs doublet, and the scalar potential invariant under the $\mathbb{Z}_{2}$ transformation $S_{i} \to -S_{i}$ (broken only by Yukawa couplings),\footnote{Alternatively, we could impose a $\mathbb{Z}_{2}\times\mathbb{Z}_{2}$ symmetry. In that case $\chi\bar{\chi} \to S_1 S_2$ would be the DM only annihilation channel. This channel is inefficient, however, because we require $S_2$ to be heavy. Consequently, we allow for a more general potential.} which we impose for convenience (in practice, values of other couplings can simply be taken to be negligible), is 
\begin{equation}
\begin{split}
  V &= \mu_{H}^{2} \abs{H}^{2} 
  + \frac{\mu_{20}^{2}}{2} S_{1}^{2}
  + \frac{\mu_{11}^{2}}{2} S_{1} S_{2} 
  + \frac{\mu_{02}^{2}}{2} S_{2}^{2}
  \\
  & 
  + \lambda_{H} \abs{H}^{4} + \lambda_{H20} \abs{H}^{2} S_{1}^{2} 
  + \lambda_{H11} \abs{H}^{2} S_{1} S_{2}
  \\
  &+ \lambda_{H02} \abs{H}^{2} S_{2}^{2}
  + \lambda_{40} S_{1}^{4} 
  + \lambda_{31} S_{1}^{3} S_{2} 
  + \lambda_{22} S_{1}^{2} S_{2}^{2} 
  \\
  &+ \lambda_{13} S_{1} S_{2}^{3} 
  + \lambda_{04} S_{2}^{4},
\end{split}
\label{eq:V:general:2:fields:H}
\end{equation}
where the numeric indices of couplings count powers of the $S_{1}$ and $S_{2}$ fields.

We consider a phase transition pattern in which at zero temperature the VEVs of $S_{1}$ and $S_{2}$ fields vanish.\footnote{Because of this and the $\mathbb{Z}_2$ symmetry, there is no mixing between the singlet scalars and the Higgs at zero temperature. For that reason, the DM direct detection signal in this model is suppressed by loops.} Hence the Higgs mass term and quartic self-interaction are related to its mass $M_h = 125.09$~GeV and the Higgs field VEV $v=246$~GeV in the usual way via
\begin{equation}
    \mu_{H}^{2} = -\frac{M_{h}^{2}}{2}\,, \quad \lambda_{H} = \frac{M_{h}^{2}}{2v^{2}}.
\end{equation}
The $S_1$ and $S_2$ fields mix in the $T=0$ vacuum, and squared masses of the mass eigenstates are given by the eigenvalues of the matrix
\begin{equation}
M_{12}^2 = \begin{pmatrix} 
\mu_{20}^2 + \lambda_{H20}v^2 & \frac{1}{2} \mu_{11}^2 + \frac{1}{2} \lambda_{H11} v^2 \\
\frac{1}{2} \mu_{11}^2 + \frac{1}{2} \lambda_{H11} v^2 & \mu_{02}^2 + \lambda_{H02}v^2 
\end{pmatrix} .
\end{equation}
The mass of $\chi$ is set purely by the bare mass parameter $M_\chi$.

Large $\lambda_{31}$, $\lambda_{13}$ and $\lambda_{H11}$ can make the potential not bounded from below, which can be compensated for by larger singlet self-couplings or Higgs portals of the singlets. We take into account the full analytical bounded-from-below (BfB) conditions for the scalar potential using the results of Ref.~\cite{Kannike:2016fmd}.

We are interested in the parameter space region where the DM freeze-out occurs before the EWPT. This puts several conditions on the field content, masses and couplings of the model. With only one singlet scalar $S_{1}$, the dominant annihilation of DM would be through the $t$-channel process $\chi \bar{\chi} \to S_{1} S_{1}$. This process is effective in the EW minimum as well, and the thermal effects we are interested in can not be realized. For that reason, we set the Yukawa coupling of $S_{1}$ to DM to zero, $y_1=0$. Consequently, we need the $S_{2}$ singlet scalar whose Yukawa coupling with $\chi$ is non-zero, and we set $M_{2} > M_{\chi}$ which forbids the annihilation $\chi \bar{\chi} \to S_{2} S_{2}$ kinematically. The DM annihilation channels in this case are discussed in Section~\ref{sec:freezeout}. Next we will study details of the EWPT.

\section{Electroweak phase transition}
\label{sec:ewpt}

The leading order high temperature thermal corrections to the potential are given by temperature dependent mass parameters
\begin{equation}
\begin{aligned}
&\mu_H(T)^2 = \mu_H^2 + c_H T^2\,, \\ &\mu_{20}(T)^2 = \mu_{20}^2 + c_{20} T^2\,,
\end{aligned}
\label{eq:thermal:mass:param}
\end{equation}
where 
\begin{equation}
\begin{aligned}
  c_{H} &= \frac{1}{48} (24 \lambda_{H} + 3 g^{\prime 2} + 9 g^{2} + 12 y_{t}^{2} + 4 \lambda_{H20}), \\ 
  c_{20} &= \frac{1}{6} (6 \lambda_{40} + 2 \lambda_{H20}).
\end{aligned}
\end{equation}
We choose the parameters such that the minimum in the $S_2$ direction is at $S_2=0$, and neglect both thermal contributions to the mass of $S_{2}$ and the contributions from any interactions that involve $S_{2}$ to other parameters, assuming that it is heavy compared to the EWPT temperature.

\begin{figure}[tb]
\begin{center}
\includegraphics{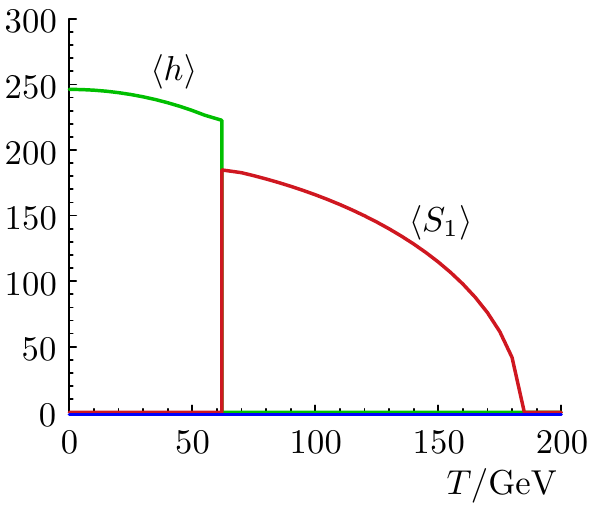}
\caption{Thermal evolution of the VEVs of fields (in GeV). DM freezes out in the minimum where $\langle S_{1} \rangle$ (red) is non-zero and $\langle h \rangle$ (green) is zero. The heavy singlet (blue) does not get a VEV.}
\label{fig:thermal:evol}
\end{center}
\end{figure}

Let us consider a phase transition pattern, where $S_{1}$ gets a VEV $\langle S_1\rangle \equiv w \lsim v$ before the EW symmetry is broken. As the temperature decreases, there is transition to the EW-breaking minimum where $\langle S_1\rangle = 0$. The thermal evolution of the VEVs for a typical example is shown in Fig.~\ref{fig:thermal:evol}. We assume that $\langle S_1\rangle = 0$ in the EW-symmetric minimum and set $\mu_{11}^2=\lambda_{31}=\lambda_{13}=0$. So, the singlet scalar do not mix, and their masses in the high temperature vacuum are\footnote{We distinguish the masses in the EW-symmetric minimum from the $T=0$ masses by tilde, $\tilde M_j$.}
\begin{equation}
\begin{aligned}
&\tilde M_1^2 = \mu_{20}^2 + 12 \lambda_{40}w^2, \\
&\tilde M_2^2 = \mu_{02}^2 + 2 \lambda_{22}w^2.
\end{aligned}
\label{eq:thermal:masses}
\end{equation}
The mass of the Higgs field is
\begin{equation}
\tilde M_h^2 = \mu_H^2 + \lambda_{H20}w^2.
\label{eq:thermal:mass:Higgs}
\end{equation}
In Eqs. \eqref{eq:thermal:masses} and \eqref{eq:thermal:mass:Higgs}, the mass parameters are the temperature-dependent expressions of Eq. \eqref{eq:thermal:mass:param}.
As described in the end of the previous section, we set $y_1=0$, so the mass of $\chi$ is $M_\chi$ also in the EW-symmetric minimum.

In order to realize the above phase transition pattern, the following conditions must be satisfied: first, the $S_{1}$ field must get a VEV before the Higgs field does, which is ensured by the conditions $\mu_{20}<0$ and
\begin{equation}
  \frac{\mu_{20}^{4}}{c_{20}^{2}} > \frac{\mu_{H}^{4}}{c_{H}^{2}}.
\end{equation}
Second, to ensure that at $T = 0$ the EW minimum is the global one, we must have
\begin{equation}
  \frac{\mu_{H}^{4}}{\lambda_{H}} > \frac{\mu_{20}^{4}}{4 \lambda_{40}}.
\end{equation}
Under these conditions, at the critical temperature $T_c$ there is a potential barrier between the minima. Then, the phase transition is of first order, unless
\begin{equation}
\frac{\mu_{20}(T)^2}{\lambda_{40}} > \frac{\mu_H(T)^2}{\lambda_{H20}}
\end{equation}
at temperatures $T\leq T_c$ before the transition happens. Hence, we need to check that the potential energy difference between the EW-symmetric and EW-breaking minima becomes sufficiently large compared to the height of the potential barrier between them, enabling formation of EW vacuum bubbles which expand and finally fill the Universe. The nucleation rate for these bubbles per unit of time and volume is given by~\cite{Linde:1981zj}
\begin{equation} \label{bubbleprob}
\Gamma(T) \simeq T^4\left(\frac{S_3(T)}{2\pi T}\right)^{3/2}\exp\left( -\frac{S_3(T)}{T} \right),
\end{equation}
where\footnote{The effect of $S_2$ on the bubble nucleation is negligible, because its displacement from zero in the bubble wall region due to the $\lambda_{H11}$ term is suppressed by $\mu_{02}^2\gg v^2$.}
\begin{equation} \label{fullS3}
S_3(T) = 4\pi \int r^2 {\rm d}r \left(\frac{1}{2}\left(\frac{{\rm d}h}{{\rm d}r}\right)^2 + \frac{1}{2}\left(\frac{{\rm d}S_1}{{\rm d}r}\right)^2 + \tilde V(T) \right)
\end{equation}
is the three-dimensional Euclidean action for an O(3)- symmetric bubble corresponding to the path in the field space which minimizes the action $S_3$. The scalar potential $\tilde V$ in~\eqref{fullS3} is defined as $\tilde V(T) = V(T) - V(T;h=0,S_1=w)$. We calculate the path which minimizes $S_3$ using the method of Ref.~\cite{Konstandin:2006nd}. Finally, the bubble nucleation temperature $T_{\rm n}$ is defined as the temperature at which the probability of producing at least one bubble per horizon volume in Hubble time is high, that is
\begin{equation}
\frac{4\pi}{3} \frac{\Gamma(T_n)}{H(T_n)^4}\simeq 1\,.
\end{equation}

\section{Dark matter freeze-out}
\label{sec:freezeout}

We assume that the DM freeze-out happens before the EWPT, $M_{\chi}/20 \gsim T_n$. As indicated by the analysis of Ref.~\cite{Vaskonen:2016yiu}, it is difficult to realize EWPT at temperatures $T_{\rm n}\lsim 50$\,GeV with the two-step transition pattern (at least for $\lambda_{40}\simeq 0.025$). So, the mass of the DM particle has to be $M_{\chi} \gsim 1$\,TeV for its freeze-out to happen before the EWPT.

\begin{figure*}[tb]
\begin{center}
\includegraphics{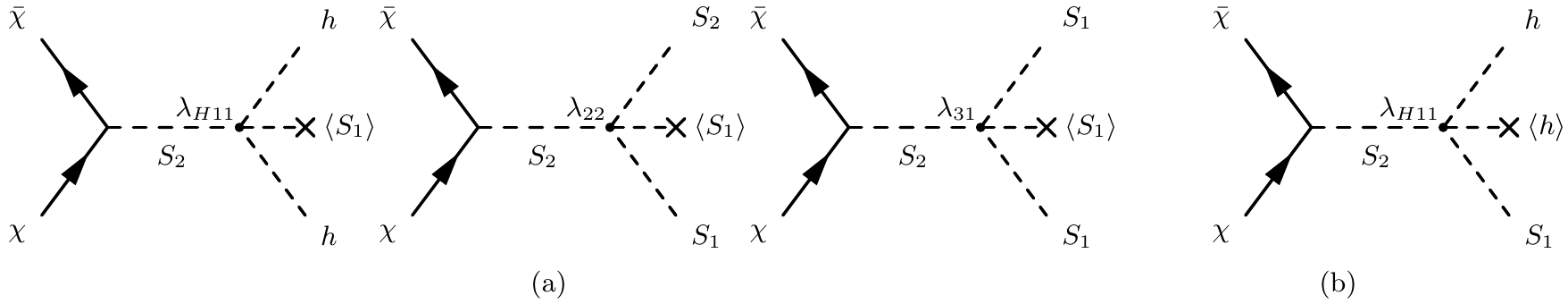}
\caption{(a) DM annihilation channels in the EW-symmetric minimum. The freeze-out is dominated by the $\chi \bar{\chi} \to h h$ channel. (b) The indirect signal of the annihilation of dark matter in the EW-breaking minimum at $T = 0$.}
\label{fig:feynman}
\end{center}
\end{figure*}

There are three $s$-channel diagrams that contribute to the DM annihilation process shown in Fig.~\ref{fig:feynman}: 
(1) $\chi \bar{\chi} \to S_{1} S_{2}$, proportional to $\lambda_{22}$, 
(2) $\chi \bar{\chi} \to S_{1} S_{1}$, proportional to $\lambda_{31}$,
(3) $\chi \bar{\chi} \to h h$, proportional to $\lambda_{H11}$, 
of which the last process arises as the dominant one. In general, the large mass of $S_2$ suppresses process (1) compared to process (2). For simplicity, we set $\lambda_{22}$ to zero as the channel is phenomenologically uninteresting.\footnote{For reasonable values of $\lambda_{22}$ the freeze-out is still dominated by the process (3) for our benchmark point.} We set $\lambda_{31}$ to zero to avoid conveying the Yukawa interaction between $\chi$ and $S_{2}$ also to $S_{1}$ via mixing between $S_{1}$ and $S_{2}$, so process (2) is altogether absent for our benchmark points shown below.
 
The freeze-out cross-section to the $hh$ final state is given by \cite{Kainulainen:2015sva}
\begin{equation}\label{eq:fo-x-section}
\sigma_{\rm fo} = \frac{1}{32\pi} \sqrt{\frac{s - 4\tilde M_h^2}{s-4 M_\chi^2}} \frac{y_2^2 \lambda_{H11}^2 w^2}{(s - \tilde M_2^2)^2 + \tilde M_2^2 \Gamma_{2}^2} \,,
\end{equation}
where $\Gamma_{2}$ is the decay width of $S_2$. Note that $\tilde M_h$ and $w$ depend on temperature, and $\tilde M_2^2 \simeq \mu_{02}^2$. The mass of $\chi$, instead, is determined by the bare mass parameter only, because the Yukawa coupling $y_1$ is set to zero. We calculate the freeze-out in the standard way  by solving the Zeldovich-Okun-Pikelner-Lee-Weinberg equation~\cite{Zeldovich:1965,Lee:1977ua}. We take into account thermal masses and VEVs by assuming that all thermal dependent properties remain constant during the freeze-out process itself.\footnote{We have checked that this gives a sufficiently good approximation for the relic abundance, as the freeze-out process happens in a very narrow temperature range.} The value of the product of the couplings $y_2 \lambda_{H11}$ is then fixed such that the observed DM abundance, $\Omega_\chi h^2 = 0.1188$~\cite{Ade:2015xua}, is obtained.

We consider a benchmark point with $\mu_{20}^2=-3820\,{\rm GeV}^2$, $\mu_{11}^{2} = 0$, $\lambda_{40}=0.025$, $\lambda_{04}=0.15$, $\lambda_{H20}=0.27$, $\lambda_{H02}=0.55$, and  $\lambda_{22}=\lambda_{13}=\lambda_{31}=0$. We pick three different values for $\mu_{02}^2$ corresponding to $M_2=3000,\,3400,\,3800$\,GeV. These parameter sets satisfy all the conditions given in Sec.~\ref{sec:ewpt}.\footnote{We calculated the renormalization group equations with the PyR@TE 2 package \cite{Lyonnet:2013dna,Lyonnet:2016xiz}. For the presented benchmark point, the Landau pole arises on the scale from $10^{4}$ to $10^{7}$\,GeV around the resonance ($y_{2} \lambda_{H11} \lesssim 3$).} In particular, the values of the quartic couplings are chosen as to satisfy the BfB conditions, and the value of $\lambda_{H20}$ is bounded from above from the requirement of a successful EWPT.
 The bubble nucleation temperature for this point is $T_n=62$\,GeV.

\begin{figure}[tb]
\begin{center}
\includegraphics{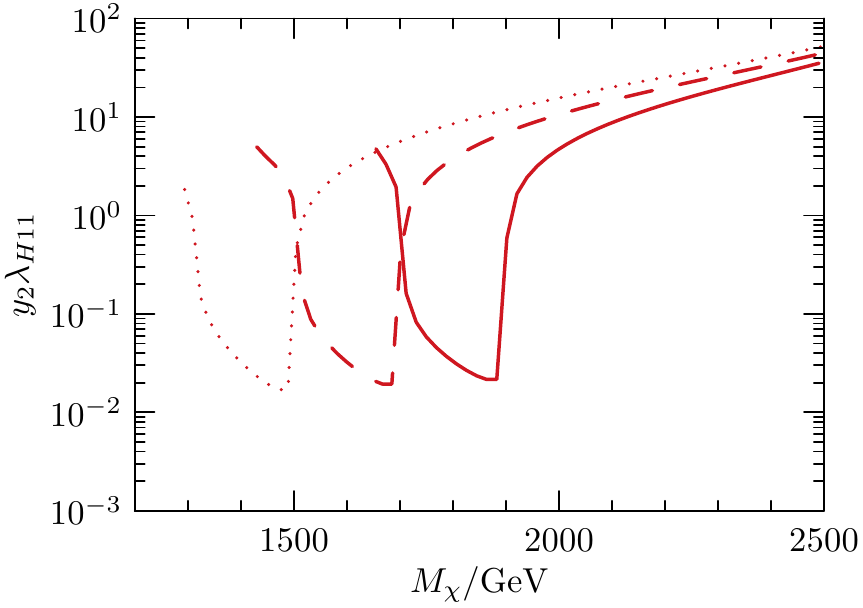}
\caption{Parameter space for the fixed benchmark sets producing the observed DM abundance. The different colors are for $M_{2} = 3000$~GeV (dotted), $3400$~GeV (dashed) and $3800$~GeV (solid).}
\label{fig:param:space}
\end{center}
\end{figure}

Having fixed the benchmark points, Fig.~\ref{fig:param:space} depicts the parameter space that produces the correct relic density for the range of $y_2\lambda_{H11}$ against $M_\chi$. The product $y_2\lambda_{H11}$ remains of order unity or less at the wide areas around the resonance, $2 M_\chi \simeq M_2$. The left end of each line is determined by the phase transition temperature or correct relic density (whichever gives a strongest bound), the right end by perturbativity.

In the end, $S_{1}$ and $S_{2}$ have to either annihilate or decay to ensure that $\chi$ is the dominant component of DM. For the benchmark points $S_2$ finally decays to $hS_1$. One should also allow for small $\mathbb{Z}_{2}$ breaking terms, such as $\mu_{H10} \, S_1 \abs{H}^{2}$, to make $S_1$ unstable as the abundance of $S_1$ is strongly constrained by direct detection. In the EW vacuum, the large $\lambda_{H11}$ and the Higgs VEV cause a large direct detection cross-section.

\section{Indirect detection}
\label{sec:detection}

The panel (b) of Fig.~\ref{fig:feynman} shows the process which produces the indirect signal in the EW vacuum, $T \simeq 0$. Assuming again small $\mathbb{Z}_{2}$ breaking term(s), the on-shell $S_1$ decays dominantly to $b\bar b + b \bar b$ via a pair of (off-shell) Higgs bosons. Thus the final state of the annihilation includes 3 pairs of $b\bar b$.

\begin{figure}[tb]
\begin{center}
\includegraphics{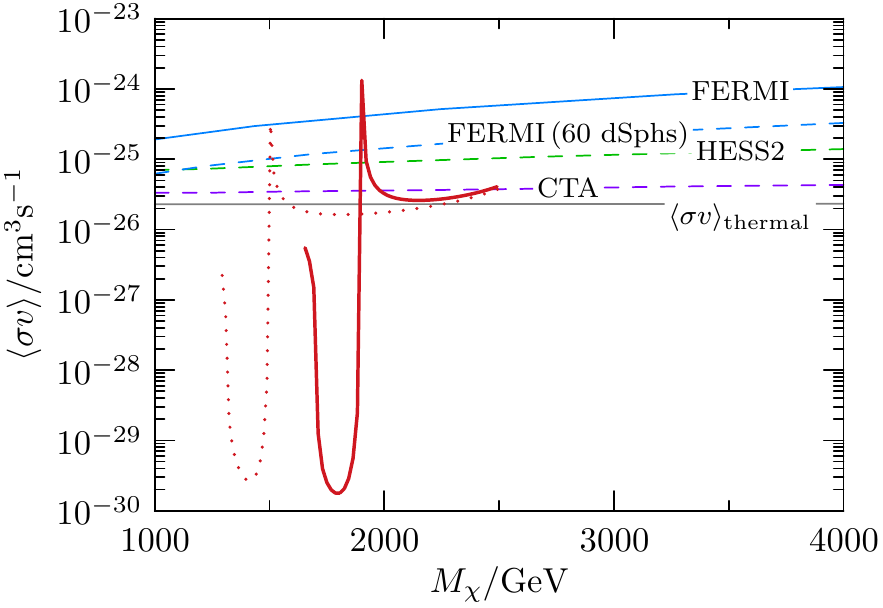}
\caption{Estimated indirect signal of the model for the benchmark point with $M_2=3000$~GeV (dotted red) and 3800~GeV (solid red), and the present constraint (solid blue) and a future estimate from dwarf spheroidal satellite galaxies (dashed blue) by FERMI~\cite{Ackermann:2015zua}, a preliminary result from the Galactic Centre by HESS2~\cite{Abramowski:2011hc} (dashed green) and an estimate from the Galactic Centre by CTA~\cite{Pierre:2014tra} (dashed purple). The thermal velocity-averaged cross section is shown in gray.}
\label{fig:indirect}
\end{center}
\end{figure}

In the non-relativistic limit the thermally averaged cross section relevant determining the DM freeze-out at $T>T_n$ is well approximated by taking $s \approx 4M_\chi^2(1+v_\chi^2)$:
\begin{equation}
\langle v\sigma_{\rm fo}\rangle \approx  \frac{1}{64\pi} \frac{y_2^2 \lambda_{H11}^2 w^2}{(4M_\chi^2(1+v_\chi^2)-\mu_{02}^2)^2+\mu_{02}^2 \Gamma_2^2}\,,
\end{equation}
where we assumed that $M_\chi\gg M_h$. Similarly, the thermally averaged cross section for the DM indirect detection (at $T\simeq0$) is given by\footnote{The factor of 2 difference in the cross sections arises because for indirect detection the final state is $S_1 h$ while for freeze-out it is $hh$.}
\begin{equation}
\langle v\sigma_{\rm indirect}\rangle \approx \frac{1}{32\pi} \frac{y_2^2 \lambda_{H11}^2 v^2}{(4M_\chi^2-\mu_{02}^2)^2+\mu_{02}^2 \Gamma_2^2}\,.
\end{equation}
Thus the cross-sections for the indirect detection and the freeze-out are, up to velocity factors, related by
\begin{equation}
\frac{\langle v\sigma_{\rm indirect}\rangle}{\langle v\sigma_{\rm fo}\rangle} \propto \frac{2v^2}{w^2}\,.
\label{eq:xsec:ratio}
\end{equation}
Fig.~\ref{fig:indirect} shows the estimate of the indirect signal for the benchmark points with $M_2=3000$~GeV and 3800~GeV and the DM mass and $y_2\lambda_{H11}$ ranges corresponding to Fig.~\ref{fig:param:space}. The cut-off of the red lines at large masses arises from perturbativity requirement.

The large changes of the indirect detection cross-section as a function of mass (the red curves in Fig.~\ref{fig:indirect}) are caused by the propagator of Eq.~(\ref{eq:fo-x-section}). The minimum of the cross-section in the left side of the red curve is caused by the fact that the freeze-out happens close to the pole of the propagator (so $y_2 \lambda_{H11}$ is small). The sudden narrow maximum of the cross-section on the right side of the large minimum is due to the pole of the indirect detection cross-section. As the freeze-out happens at typical velocities $v \simeq 0.1$ and the indirect detection at $v \simeq 10^{-3}$, the minimum and maximum do not cancel each other and appear at different values of $M_\chi$. These effects are usual to indirect detection. We emphasise that, even though the main effect that causes the difference between the red lines and the na\"ive thermal cross section shown by the solid gray line is due to the resonance, the thermal effects enhance the indirect detection cross section by the factor given by Eq. \eqref{eq:xsec:ratio}. 

A set of current and future constraints are included: the present constraint and a future estimate from dwarf spheroidal satellite galaxies by the Fermi LAT~\cite{Ackermann:2015zua}, a preliminary result from the Galactic Centre by the HESSII~\cite{Abramowski:2011hc} and a future estimate from the Galactic Centre by the CTA experiment~\cite{Pierre:2014tra}. The constraints are for the direct $2 \, \mathrm{DM} \to b \bar b$ final state. In our case the $b$ final states originating from the $h$ and $S_1$ decays are boosted. To estimate the constraints, we compared the $\gamma$-ray signals from the $b$ and $h$ final states using the PPPC4DMID toolkit~\cite{Cirelli:2010xx}. The comparison shows that the $\gamma$-ray signals are very similar, only a $\sim$200~GeV shift of the mass of $\chi$ has to be introduced to have a good match. Using the comparison, one can  convert the constraints on the $b$ final state to the constraints on the $h$ final state. The $h$ final state spectrum has some extra features compared to the $b$ one, e.g. a small tip at the higher energy end, due to the other decay channels of $h$. Naturally, those features make the $h$ final state more visible over the power-law like astrophysical backgrounds. In our approach we neglect those features; in this sense, our estimate is on the conservative side.

\section{Gravitational wave signal}
\label{sec:gw}

As shown in Sec.~\ref{sec:ewpt} the EWPT is typically of first order for the model parameters which enable DM freeze-out in EW-symmetric vacuum. The first order phase transition proceeds via nucleation of bubbles which expand and eventually collide. The bubble collisions and the motion of the plasma after the collision source GW background (for a recent review see Ref.~\cite{Weir:2017wfa}). The spectrum of this background is determined by the following three parameters: the ratio of released vacuum energy in the transition to that of the radiation bath, 
\begin{equation}
\alpha = \frac{1}{\rho_\gamma(T_n)}\left(\Delta V - \frac{T_n}{4}\Delta\frac{{\rm d}V}{{\rm d}T} \right)\bigg|_{T=T_n}\,,
\end{equation}
the inverse duration of the phase transition,
\begin{equation}
\beta = H(T_n) T_n \frac{{\rm d}}{{\rm d} T} \frac{S_3(T)}{T}\bigg|_{T=T_n}\,,
\end{equation}
which can be easily calculated, once the bubble nucleation temperature $T_n$ is known, and the bubble wall velocity $\xi_w$, which we consider a constant for simplicity. 

\begin{figure}[tb]
\begin{center}
\includegraphics{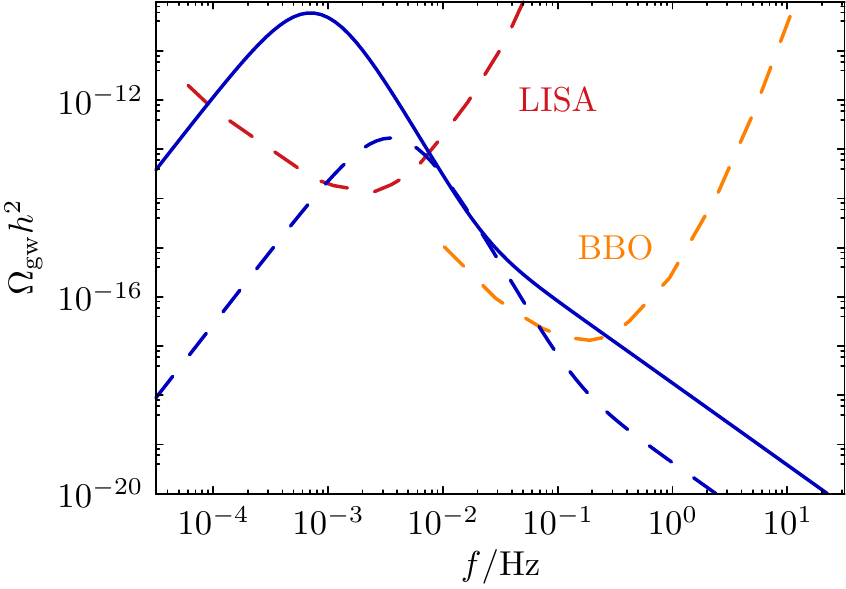}
\caption{The GW background arising from the first order EWPT corresponding to the benchmark point given in Sec.~\ref{sec:freezeout} for $\xi_w=0.5$ (solid blue) and $\xi_w=0.1$ (dashed blue). The dashed red and orange lines show the expected sensitivities of the future GW interferometers~\cite{Thrane:2013oya}.}
\label{fig:grav}
\end{center}
\end{figure}

We have checked that $\alpha<\alpha_\infty$~\cite{Bodeker:2009qy,Espinosa:2010hh} for the benchmark point considered in Sec.~\ref{sec:freezeout}, so the bubble wall does not runaway. Using the formulae given in Ref.~\cite{Caprini:2015zlo} we calculate the spectrum of the GW background from the first order EWPT, $\Omega_{\rm GW} h^2(f)$. The result, for two different values of $\xi_w$, is shown in Fig.~\ref{fig:grav}. Shown is an extreme case, and typically the signal is much weaker.

In our model, the strength of the EWPT is not directly related to the parameters which determine the DM freeze-out. The stronger the transition is, however, the later it typically happens, thus enabling DM freeze-out before EWPT for lower DM masses. In that way, we expect a strong GW background in the case of light DM, $M_\chi \simeq 1$\,TeV, and its freeze-out happens before EWPT.

\section{Conclusions}
\label{sec:conclusions}

We presented a singlet fermion DM model, where the freeze-out happens in a high-temperature minimum. In addition, we required two real singlet scalars, the light one to produce the thermal phase transition and the heavy one to act as a mediator between DM and the SM. Due to thermal evolution, the light singlet scalar temporarily obtains a VEV, which opens efficient annihilation channels for the freeze-out of DM. The DM relic density, then, differs from a na\"ive calculation in the EW breaking minimum.  

We studied theoretical and experimental constraints on the model. An indirect signal from DM annihilation to the $S_1 h$ final state is present in the EW vacuum, and the cross-section for this process is different by factor $\sim 1/2\, (v/w)^2$ from the one which determines DM freeze-out. While in the case of `forbidden' annihilation channels~\cite{DAgnolo:2015ujb, Griest:1990kh}, for example, the constraints from indirect detection can be evaded, we can realize the case $v/w > 1$. Our results demonstrate that the thermal effects on DM freeze-out can have important observational ramifications.

The first order two-step EWPT in the model produces a stochastic GW background potentially detectable at upcoming experiments. Its properties are not directly related the DM freeze-out related parameters of the model, but the GW signal is strongest for $M_\chi \simeq 1$\,TeV, corresponding roughly to the minimal DM mass for which the DM freeze-out in our model can happen before the EWPT. The lighter the DM particle is, the later its freeze-out must happen, and for the freeze-out to still happen before the EWPT the EWPT has to be delayed, which strengthens the transition.

\vspace{1.em}

\begin{acknowledgments}
This work was supported by the Estonian Research Council grant PUT799 and PUT808, the grant IUT23-6 of the Estonian Ministry of Education and Research, and by the EU through the ERDF CoE program project TK133. AH thanks the Horizon 2020 programme as this project has received funding from the programme under the Marie Sklodowska-Curie grant agreement No 661103.
\end{acknowledgments}

\end{fmffile}

\bibliography{foamin}

\end{document}